# One-Dimensional Crystallographic Etching of Few-Layer WS$_2$

*Shisheng Li,[1]\* Yung-Chang Lin,[2,3] Yiling Chiew,[3] Yunyun Dai,[4] Zixuan Ning,[4] Hideaki Nakajima,[5] Hong En Lim,[6] Jing Wu,[7] Yasuhisa Naito,[8] Toshiya Okazaki,[5] Zhipei Sun,[4] Kazu Suenaga,[2,3] Yoshiki Sakuma,[9] Kazuhito Tsukagoshi[1] and Takaaki Taniguchi[1]*

1 Research Center for Materials Nanoarchitectonics (MANA), National Institute for Materials Science (NIMS), Tsukuba 305-0044, Japan

2 Nanomaterials Research Institute, National Institute of Advanced Industrial Science and Technology, AIST Central 5, Tsukuba 305-8565, Japan

3 The Institute of Scientific and Industrial Research (ISIR-SANKEN), Osaka University, Osaka 567-0047, Japan

4 QTF Centre of Excellence, Department of Electronics and Nanoengineering, Aalto University, Espoo 02150, Finland

5 Nano Carbon Device Research Center, National Institute of Advanced Industrial Science and Technology, AIST Central 5, Tsukuba 305-8564, Japan

6 Department of Chemistry, Graduate School of Science and Engineering, Saitama University, Saitama 338-8570, Japan

7 Institute of Materials Research and Engineering (IMRE), Agency for Science, Technology and Research (A*STAR), 138634, Singapore

8 Nanosystem Research Institute, National Institute of Advanced Industrial Science and Technology (AIST), Tsukuba 305-8562, Japan

9 Research Center for Electronic and Optical Materials, National Institute for Materials Science (NIMS), Tsukuba 305-0044, Japan

\*Correspondence to: li.shisheng@nims.go.jp

**Keywords:** transition metal dichalcogenides, etching, one-dimensional nanotrenches, photoluminescence, second harmonic generation.

**Abstract:** Layer number-dependent band structures and symmetry are vital for the electrical and optical characteristics of two-dimensional (2D) transition metal dichalcogenides (TMDCs). Harvesting 2D TMDCs with tunable thickness and properties can be achieved through top-down etching and bottom-up growth strategies. In this study, we report a pioneering technique that utilizes the migration of in-situ generated Na-W-S-O droplets to etch out one-dimensional (1D) nanotrenches in few-layer $WS_2$. 1D $WS_2$ nanotrenches were successfully fabricated on the optically inert bilayer $WS_2$, showing pronounced photoluminescence and second harmonic generation signals. Additionally, we demonstrate the modulation of inkjet-printed $Na_2WO_4$-$Na_2SO_4$ particles to switch between the etching and growth modes by manipulating the sulfur supply. This versatile approach enables the creation of 1D nanochannels on 2D TMDCs. Our research presents exciting prospects for the top-down and bottom-up fabrication of 1D-2D mixed-dimensional TMDC nanostructures, expanding their use for photonic and optoelectronic applications.

**1. Introduction**

Etching plays a vital role in semiconductor manufacturing as it removes unwanted materials through chemical and physical processes. This technique, in conjunction with advanced lithography and thin-film deposition methods, enables the fabrication of intricate and precise circuit designs essential for high-performance electronics. However, as downsizing of three-dimensional (3D) silicon-based metal-oxide-semiconductor field-effect transistors (FETs) is approaching their physical limits, further enhancements in performance are anticipated to reach a plateau. In contrast, two-dimensional (2D) transition metal dichalcogenides (TMDCs) have emerged as a promising alternative to silicon in the post-Moore era, owing to their atomically thin layered structure, defect-free surfaces, and exceptional electrical and optical properties.

Etching is a critical process in the utilization of 2D TMDCs. A comprehensive understanding of the etching mechanism and developing diverse etching techniques are essential for expanding their applications across various fields. [1,2] For example, lithography-assisted patterning and plasma etching are indispensable for the scalable integration of wafer-size 2D TMDC-based functional devices, including FETs and light-emitting diodes (LEDs). [3-6] Controlled layer-by-layer etching of 2D TMDCs can be achieved through the use of ion beams with moderate power, allowing precise modulation of their band structures. [7,8] Additionally, lasers can be employed to thin and pattern atomically thin sheets, resulting in the formation of sub-micrometer-size holes (0D) and complex 1D and 2D patterns on 2D TMDCs. [9-11] These laser-induced etching spots serve as ideal nucleation sites for the growth of top metallic 2D TMDCs, facilitating the realization of high-performance van der Waals contacts. [12,13] Furthermore, mild etching conditions applied to these spots result in the formation of large holes in the basal plane of 2D TMDCs, which have been utilized for the epitaxial growth of other TMDCs, leading to the formation of monolayer mosaic heterostrucutres. [11,14]

The natural tendency for thermodynamically favored etching in the basal plane of most 2D materials with a hexagonal crystal structure is limited to the formation of micrometer-scale 2D hexagonal and triangular holes. [1,11,14] Therefore, it is crucial to develop novel methods for achieving 1D crystallographic etching in 2D materials. Encouragingly, the utilization of Fe, Co, Ni, or $SiO_2$ nanoparticles (NPs) had demonstrated successful 1D crystallographic etching of few-layer graphene under reducing atmospheres at high temperature. This process leads to the formation of 1D nanotrenches (NTs) with depths of a few atomic layers within the basal plane of graphene. [15-21] It is noteworthy to mention that these catalysts are also suitable for the growth of 1D carbon nanotubes and graphene nanoribbons (NRs). [22-25] The presence or absence of a carbon source during the high-temperature process governs the reversibility of these reactions. In the case of TMDCs, vapor-liquid-solid (VLS) epitaxy of 1D $MoS_2$ NRs on monolayer $MoS_2$ film has been achieved by precipitating solid $MoS_2$ from S-saturated Na-Mo-O or Na-Mo-Ni-O droplets.[26,27] This unique bottom-up growth process (Snake mode) enables the fabrication of mixed dimensional 1D-2D TMDCs (**Fig. 1a**). Conversely, the reverse 1D crystallographic etching process (Pac-Man mode) provides an alternative top-down strategy for preparing such mixed-dimensional TMDCs (**Fig. 1b**). The presence of these 1D NRs and NTs significantly impacts the electrical and optical properties of TMDCs due to the layer-dependent band structures and symmetry. However, the underlying mechanism and methods for controlling the switch between 1D crystallographic etching and growth of TMDCs remain unclear. Since this approach could potentially serve as the ultimate method for achieving both top-down etching and bottom-up growth of the same materials using a single droplet, it thus requires urgent elucidation. Recently, 1D crystallographic etching has been observed in KOH-decorated thick $MoS_2$ flakes under ambient oxidizing conditions.[28] However, the method is limited to thick $MoS_2$ flakes, which does not offer the ability to modulate the bandgap and optical properties of 1D-2D mixed-dimensional nanostructures.

In this study, we present a novel approach to chemically etch molten-salt chemical vapor deposition (CVD)-grown $WS_2$ flakes in a mild reducing environment devoid of sulfur. The etching process utilizes in-situ generated Na-W-S-O droplets on the $WS_2$ surface, resulting in the formation of unique 1D NTs with distinct crystallographic patterns in the few-layer (≥2) regions. The 1D NTs (monolayer) embedded within the optically inert bilayer $WS_2$ exhibit strong photoluminescence (PL) and second harmonic generation (SHG) signals due to the transition from an indirect to a direct bandgap and from a centrosymmetric to a non-centrosymmetric state. Comprehensive chemical analysis confirms that the droplets consist of Na, W, S, and O. We elucidate the crucial roles and evolutionary dynamics of Na, W, S, and O during the 1D crystallographic etching process. Notably, we demonstrate that the S plays a decisive role in determining whether the inkjet-printed Na-W-S-O ($Na_2WO_4$-$Na_2SO_4$) droplets would proceed in either the etching or growth mode. The integration of top-down etching and bottom-up growth strategies using molten salt NPs holds great promise for engineering 1D-2D mixed-dimensional TMDC nanostructures with tailored functionalities.

## 2. Results and Discussions
### 1D crystallographic etching of few-layer WS$_2$

To investigate the phenomenon of 1D crystallographic etching in 2D TMDCs, we employ molten salt CVD-grown WS$_2$ flakes. The molten salt-CVD method involves the direct sulfurization or selenization of molten salts (e.g., Na$_2$MoO$_4$, Na$_2$WO$_4$, etc.) to achieve the growth of both 1D and 2D TMDC monolayers in a VLS mode. [29-32] As depicted in **Fig. 2a** and **Supplementary Fig. 1a**, the as-grown WS$_2$ flakes exhibit large grain sizes reaching several hundred micrometers. Optical contrast analysis readily allows the identification of monolayer (+1), bilayer (+2), trilayer (+3), and even thick-layer (≥4) regions. [33] These WS$_2$ flakes, deposited on SiO$_2$/Si substrates, are subsequently loaded into the same CVD system equipped with a clean quartz tube. The etching experiments are primarily conducted at a temperature of 600 °C in a mild reducing environment composed of Ar-(0.5$_{vol.}$%) H$_2$ forming gas without sulfur. It is observed that at other temperatures, etching efficiency is poor at 550 °C, while severe degradation of WS$_2$ occurs at high etching temperatures above 650 °C (**Supplementary Fig. 2**). At the optimal etching temperature of 600 °C, a high yield of 1D NTs and minimal degradation of WS$_2$ are achieved simultaneously. Raman spectra and mapping results demonstrate that the characteristic peaks of WS$_2$ remain largely unchanged in the as-etched WS$_2$ flakes (**Supplementary Fig. 3**). [34]

**Fig. 2b**, **2c,** and **Supplementary Fig. 1b-1f** depict typical optical and SEM images of WS$_2$ flakes after etching at 600 °C. High-density 1D NTs terminated with NPs are observed in the bilayer and trilayer regions of the as-etched WS$_2$ flakes. In **Fig. 2c**, two NPs etch from the trilayer (+3) to the bilayer (+2) regions of the WS$_2$ flake, leaving behind two 1D NTs on the topmost layers. Interestingly, we observe that the NPs do not cross over each other but instead turn at angles of 120° when encountering other 1D NTs (**Fig. 2c**). This observation can be explained by the epitaxial AA and AB stacking of hexagonal bilayer $WS_2$, which leads to unique angles of 0°, 60°, and 120° between two 1D NTs. This finding supports the occurrence of 1D crystallographic etching on the topmost $WS_2$ layer.[26,28] SEM observations of over three hundred NPs and as-etched 1D NTs confirm that the etching is limited to the topmost layers of few-layer $WS_2$ under our mild etching conditions. The NPs are unable to simultaneously etch through two or thicker layers, likely due to the high thermodynamic barrier. Additionally, we observe that 1D crystallographic etching does not happen in monolayer $WS_2$ supported by $SiO_2$/Si substrates, and only 2D etching with triangle holes is observed (**Supplementary Fig. 4**). This provides further evidence that the bottom $WS_2$ layer plays a crucial role in guiding the movement of NPs and inducing the 1D crystallographic etching of the top layers.

Unlike the bottom-up approach (VLS epitaxy) of 1D $MoS_2$ NRs on 2D $MoS_2$ film, the 1D crystallographic etching presents a distinct top-down strategy for creating 1D NTs in 2D TMDCs, as illustrated in **Fig. 2d**. Interestingly, the 1D crystallographic etching in the same direction can also result in the formation of a comb-like structure comprising high-density parallel 1D NTs and 1D NRs on the topmost $WS_2$ layer (**Supplementary Fig. 1f**). This unique 1D-2D mix-dimensional TMDC-based nanobarcode holds potential for labeling and detection purposes.

AFM and SEM characterizations confirm that the NPs exhibit a dynamic molten state during the high-temperature etching process. **Fig. 2e** and **2f** depict the AFM topography image and height profiles of the 1D NTs and NPs on an etched $WS_2$ flake. Analysis of the height profiles reveals that the 1D NTs have a depth of ~0.7 nm, corresponding to the thickness of monolayer $WS_2$. On the other hand, the NPs exhibit a lateral size of ~200 nm and a height of ~50 nm. The lateral sizes of the NPs are slightly larger than the widths of the 1D NTs. Intriguingly, two adjacent NPs can merge to form a large NP with an elongated shape along the etching direction.

As shown in **Fig. 2g** and **2h**, two parallel 1D NTs with widths of ~110 nm and ~40 nm merge into a broader, ~130 nm-wide 1D NT due to the merging of neighboring NPs during the etching process. The dynamic evolution of NPs is further supported by the changing width of 1D NTs throughout the etching process. **Fig. 2i** demonstrates that a 1D NT widens from an initial width of ~60 nm to a final width of ~100 nm. This change in width can be attributed to the increased size of the NPs, which incorporate W and S atoms from the underlying $WS_2$ layer.

At the optimal etching temperature of 600 °C, the size of NPs and the width of 1D NTs can be controlled by adjusting the concentration of $Na_2WO_4$ used for the molten salt CVD growth of $WS_2$ flakes. The $WS_2$ flakes grown with a 20 mM $Na_2WO_4$ aqueous solution exhibit as-etched 1D NTs with a minimum width of ~6 nm (**Fig. 2j**) and a maximum width of ~98 nm (**Supplementary Fig. 5a**). Statistical analysis reveals that out of 335 1D NTs, their width distribution ranges from ~6 nm to 98 nm, with an average width of ~41 nm. Additionally, 238 NPs exhibit a size distribution of ~13 nm to 124 nm, with an average size of ~59 nm (**Fig. 2k**). On the other hand, for $WS_2$ flakes grown with a 30 mM $Na_2WO_4$ aqueous solution, SEM observation shows that the 275 as-etched 1D NTs have a width distribution of ~20 nm to 138 nm, with an average width of ~56 nm. Furthermore, the 395 NPs demonstrate a size distribution of ~23 nm to 155 nm, with an average size of ~71 nm (**Supplementary Fig.5b and 5c**). These results indicate that a higher concentration of $Na_2WO_4$ aqueous solution for $WS_2$ growth led to the formation of larger NPs and wider 1D NTs.

**Structure and optical properties of bilayer $WS_2$ with 1D NTs**

To study the morphology, atomic structure, and integrity of the as-etched $WS_2$ flakes with 1D $WS_2$ NTs, we utilized a scanning transmission electron microscope (STEM) for atomic resolution characterization. In **Fig. 3a**, two parallel 1D NTs with similar widths of ~50 nm are observed in the bilayer region of an as-etched $WS_2$ flake. **Fig. 3b** and **3c** present the atomic-resolution annular dark field (ADF) image and its corresponding atomic model, respectively, illustrating the edge of a 1D NT in bilayer WS$_2$. The image clearly reveals the hexagonal lattice of monolayer WS$_2$ in the 1D NT region, while the bilayer region exhibits a centrosymmetric AB stacking. Both the monolayer and bilayer regions of the as-etched WS$_2$ flake exhibit high crystallinity, indicating minimal impact on the WS$_2$ basal plane in our mild etching environment. The STEM images also demonstrate that the widths of 1D NTs vary continuously, and the edges of 1D NTs appear rough with a mixture of W and S atoms. This indicates that the morphology of NPs undergoes constant changes during the dynamic etching process. Although the 1D NTs with rough edges show slight deviations from the straight lines, we can still deduce that the 1D crystallographic etching occurs along the armchair direction, as depicted in **Fig. 3b** and **Fig. 3c**.

The optical properties of group-VI 2D TMDCs are primarily influenced by their thickness and symmetry. Monolayers of 2D TMDCs exhibit a direct bandgap and significantly stronger PL compared to their few-layer and bulk counterparts, which have an indirect bandgap. Moreover, odd-layer 2D TMDCs with non-centrosymmetric symmetry exhibit strong SHG intensity, while even-layer 2D TMDCs, which possess a centrosymmetric center, experience SHG quenching. [35] In **Fig. 3d** and **3e,** we present the optical and corresponding fluorescence images of an as-etched WS$_2$ flake with 1D NTs in the bilayer region. As depicted in **Fig. 3e**, the 1D NTs in the bilayer region exhibit bright fluorescence similar to the monolayer region. Conversely, the fluorescence signal is quenched in other bilayer regions due to their indirect bandgap. This observation is further supported by the PL spectra collected from 1D NTs, monolayer, and bilayer regions of the as-etched WS$_2$ flakes (**Fig. 3f**). The 1D NTs, being monolayer in nature, exhibit a strong PL emission peak at ~1.94 eV, in consistent with other monolayers. Furthermore, the absence of noticeable defect peaks at lower photon energy confirms the high crystallinity of the as-etched WS$_2$ flakes.

To investigate the optical nonlinearity of the as-etched WS$_2$ flakes embedded with 1D NTs in the bilayer regions, we conducted SHG mapping. **Fig. 3g** shows the optical image of the as-etched WS$_2$, while Fig. **3h** displays the corresponding SHG intensity map. In the SHG intensity map, no observable SHG signal is detected in the bilayer regions of WS$_2$, indicating the presence of 2H (AB) stacking with inversion symmetry in the bilayer structure. [36] However, a significant SHG signal emerges specifically at the 1D NTs, which can be attributed to the breaking of inversion symmetry in the monolayer WS$_2$ regions. The spectra collected from the monolayer, bilayer, and 1D NT regions are presented in **Fig. 3i**, further supporting the distinctive SHG characteristics of these regions.

**The mechanism of 1D crystallographic etching**

*Functions of sodium in 1D crystallographic etching*

To gain insights into the mechanism of 1D crystallographic etching, it is crucial to investigate the chemical composition of the NPs. For this purpose, energy dispersive X-ray spectroscopy (EDX) maps and spectra were collected from the as-etched WS$_2$ flakes and NPs on SiO$_2$/Si substrate (**Fig. 4a-c**). **Fig. 4a** shows a representative SEM image of an as-etched WS$_2$ flake with NPs and 1D NTs in the bilayer region. **Fig. 4b** presents the EDX map of Na, focusing on the area outlined in **Fig. 4a**. The map reveals the accumulation of Na within the NPs, indicating a non-uniform distribution on the surface. To further analyze the composition, detailed EDX spectra were collected from two random NPs. The spectra exhibit signals corresponding to Na, W, S, O, Si, and C (**Fig. 4c**). By excluding the Si signal from the SiO$_2$/Si substrate and the C signal from the environmental adsorbates, we deduce that the NPs predominantly consist of Na, W, S, and O. The presence of O is further confirmed by EELS spectrum of the residual NPs on WS$_2$ flake (**Supplementary Fig. 6**). Considering the similar composition of Na-Mo-S-O melts in VLS growth of 1D MoS$_2$ NRs, we propose that Na plays a similar role in the formation and stabilization of Na-W-S-O droplets during the high-temperature 1D crystallographic etching process. Previous studies have shown that the catalytic Na atoms from soda-lime glass, when attached to the edge of MoS$_2$, can significantly lower the energy barrier for incorporating Mo and S atoms into the MoS$_2$ lattice during growth, as compared to growth without Na atoms. [37]

Conversely, the presence of Na atoms also lowers the barrier for etching the $MoS_2$ lattice. Therefore, the abundance of Na atoms within the Na-W-S-O NPs, which are in direct contact with the top $WS_2$ layer, promotes preferential 1D crystallographic etching.

The critical role of Na in the 1D crystallographic etching is not only confirmed by its high concentration within the Na-W-S-O droplets but also by its spatial distribution in the as-grown $WS_2$ flakes and subsequential etching process. **Fig. 4d** and **4e** show an optical image and the corresponding time of flight - secondary-ion mass spectrometry (TOF-SIMS) map of $Na^+$ from the $WS_2$ flakes grown with 30 mM $Na_2WO_4$ aqueous solution. A more intense $Na^+$ signal (bright yellow color) is observed in the thick-layer $WS_2$ region (**Fig. 4e**). This can be attributed to the $WS_2$ flakes being grown from Na-W-S-O melts in a VLS mode. After the solid $WS_2$ layers have precipitated, the $Na^+$ ions become more concentrated within the remaining melts, especially in the later-grown thick-layer $WS_2$ regions (**Supplementary Fig. 7**). The layer-dependent spatial distribution of $Na^+$ leads to the in-situ formation of high-density Na-W-S-O droplets in the thick-layer regions and low-density droplets in the bilayer and monolayer regions (**Supplementary Fig. 8**). Consequently, a density gradient of 1D NTs is observed from the thick-layer to the thin-layer regions of $WS_2$ flakes (**Fig. 4f** and **Supplementary Fig. 8**). In control etching experiments, the as-grown $WS_2$ flakes are wet-transferred onto clean $SiO_2$/Si substrates using DI water. Due to the removal of residual Na-W-S-O on the $WS_2$ flakes, the transferred $WS_2$ flakes can withstand a high etching temperature of 600 °C, and only small triangle holes are observed after etching at 650 °C (**Supplementary Fig. 9**). This further demonstrates the significance of Na in the 1D crystallographic etching of $WS_2$ flakes.

*Sulfur switch for 1D crystallographic etching and growth*

In addition to the significant role of Na in the formation of Na-W-S-O droplets, we have discovered that the S concentration within the droplets is the main driving force for the 1D crystallographic etching and growth of $WS_2$. The inset of **Fig. 4c** indicates that the Na-W-S-O NPs have an S/W atomic ratio of less than <0.1, which is much lower than the as-grown $WS_2$ flakes. Further analysis using fine X-ray photoelectron spectroscopy (XPS) spectra of S$2p$ and W$4f$ reveals a significantly higher S/W atomic ratio of ~1.90 (**Fig. 5a** and **5b**). This steep S gradient is established from the bottom WS$_2$ layer to the Na-W-S-O droplets and the sulfur-free atmosphere. The Na-W-S-O droplet acts as a "pump", extracting W and S atoms from the bottom WS$_2$ layer. The digested S atoms then escape from the droplets as H$_2$S vapor in the mild reducing Ar-H$_2$ atmosphere. This process creates a sulfur-deficiency environment within the droplets and promotes the anisotropic etching of WS$_2$. In contrast, during the VLS growth of 1D MoS$_2$ NRs using Na-Mo-S-O droplets, a much higher S/Mo atomic ratio (>2) is observed, indicating an oversaturation of S in the Na-Mo-S-O droplets (**Supplementary Fig. 6f** of Ref. [26]).

To investigate the deterministic factor in switching between reversible etching and growth reactions, we conducted experiments using intentionally deposited mixed Na$_2$WO$_4$-Na$_2$SO$_4$ particles to simulate the in-situ generated Na-W-S-O nanoparticles. **Fig. 6a** illustrates an array of Na$_2$WO$_4$-Na$_2$SO$_4$ particles on a Na-free WS$_2$ flake by printing the mixed molten-salt solution. By carefully adjusting the reaction conditions, particularly the sulfur supply, the as-deposited particles could switch between 1D etching in a sulfur-free condition and 1D VLS growth in a sulfur-sufficient condition (**Fig. 6b-d**). **Fig. 6b** shows one active particle in the bilayer region and three inactive particles in the monolayer region of a WS$_2$ flake after etching at 700 °C without S. The active particle etches the top layer, leaving behind a monolayer trench with bright fluorescence in the bilayer region (**Fig. 6c**). Conversely, 1D WS$_2$ ribbons can be grown from the Na$_2$WO$_4$-Na$_2$SO$_4$ particles at 800 °C with sufficient S supply (**Fig. 6d**). Further analysis using EDX spectra reveals a higher S concentration in the Na-W-S-O particle for 1D growth compared to the particle used for 1D etching (**Supplementary Fig. 10**). Therefore, we can conclude that S is the deterministic factor in switching the Na$_2$WO$_4$-Na$_2$SO$_4$ particles between 1D crystallographic etching (Pac-Man mode) and 1D VLS growth of TMDCs (Snake mode), respectively. Currently, the yield of active particles for 1D etching and growth is still relatively low. However, future engineering efforts focusing on the chemical composition and size of the particles could lead to the fabrication of 1D TMDC nanochannels with higher yields and controlled widths.

*Evolution of tungsten during 1D crystallographic etching*

In the Na-W-S-O droplets, both W and S atoms are incorporated during the etching process of the bottom $WS_2$ layer (Inset of **Fig. 4c**). However, the higher composition of W in the droplets suggests that W has a slower escaping rate compared to S. As a result, the residual W contributes significantly to the growth of droplets and the width of derived 1D NTs. For example, in **Fig. 2e**, a 1D $WS_2$ NT with a length of ~7 µm increases its width from an initial width of ~60 nm to a final width of ~100 nm, with an increasing rate of ~5.7 nm/µm ($\delta(width)/\delta(length)$). It is worth noting that the width of 1D $MoS_2$ NRs grown through the sulfurization of Na-Mo-S-O droplets exhibits a shrinking rate of ~12.9 nm/µm ($\delta(width)/\delta(length)$) due to the continuous consumption of Mo atoms from the Na-Mo-S-O droplets (**Supplementary Fig. 7e** of Ref. [26]). Based on the similar increasing rate of 1D $WS_2$ NTs and shrinking rate of 1D $MoS_2$ NRs, it can be inferred that the gain and loss of transition metals *X* (*X* = Mo and W) play a crucial role in determining the size evolution of Na-*X*-S-O droplets, and consequently, the width of 1D NTs and NRs, respectively.

*Role of oxygen in 1D crystallographic etching*

Oxygen is a crucial component in the 1D crystallographic etching of the few-layer $WS_2$. It exists in two primary forms. Firstly, it occurs as substitutional O occupying S vacancies within the $WS_2$ lattice. This is confirmed by the presence of $WO_x$-related W*4f* peaks at 35.60 eV and 37.75 eV, [38] distinct from the dominant peaks at 32.61 eV and 34.76 eV, corresponding to the W$4f_{7/2}$ and W$4f_{5/2}$ levels of $WS_2$, respectively (**Fig. 5a**). Additionally, a weak O*1s* peak at 530.15 eV provides further evidence of $WO_x$ formation (**Fig. 5c**). [39] These substitutional O atoms in $WS_2$ have been shown to facilitate the chemical etching of 2D TMDCs. [40] The second form of O is residual $O_2$ originating from the ambient condition, which is difficult to remove from the reaction chamber even with a rotary pump. [41] Due to the presence of substitutional and environmental O sources, W atoms within the Na-W-S-O droplets can escape in the form of $WO_2(OH)_2$. This compound has a low melting point of 100 °C and has been utilized as an ideal precursor for synthesizing high-quality monolayer $WS_2$. [42] Moreover, recent studies have demonstrated the possibility of 1D crystallographic etching of thick $MoS_2$ flakes through oxidization of KOH-decorated samples in ambient conditions. In this case, S atoms within the K-Mo-S-O droplets escape as $SO_2$. [28] Hence, O plays a synergistic role in the evolution of transition metals and S within the droplets during the 1D crystallographic etching process.

Based on the analysis, one possible reaction for the 1D crystallographic etching can be proposed:

$$WS_{2-x}O_{x(s)} + Na_2S_{(s)} + H_{2(g)} + O_{2(g)} \rightarrow Na_2WO_{4(s)} + WO_2(OH)_{2(g)} + H_2S_{(g)}$$

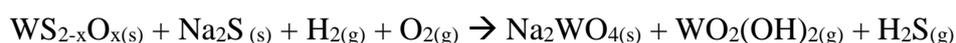

In this equation, $WS_{2-x}O_x$ represents the as-grown $WS_2$ flakes with substitutional O, $Na_2S$ represents residual Na, [43] and $O_2$ is derived from the residual air in the reaction chamber. The reaction involves the formation of $Na_2WO_4$ as a solid product, the release of $WO_2(OH)_2$ vapor, and the liberation of $H_2S$ gas. It is important to note that the proposed reaction is a simplified representation and may not capture all the intricate details of the etching process. The actual mechanisms and species involved can be more complex and may depend on various factors, including temperature, reaction conditions, and the specific properties of the starting materials.

## 3. Conclusions

The in-situ generation of sizeable Na-W-S-O droplets on top of the molten-salt CVD-grown $WS_2$ flakes results in the unique 1D crystallographic etching and leaving 1D NTs in the few-layer regions in a sulfur-free and mildly reducing environment. This 1D crystallographic etching approach represents an alternative top-down strategy, complementing the bottom-up 1D VLS growth strategy, for fabricating 1D-2D mixed-dimensional homo/heterojunctions. By etching optically inert bilayer $WS_2$, we successfully fabricated 1D $WS_2$ nanotrenches with pronounced photoluminescence and non-linear second harmonic generation signals. With the advanced inkjet printing technique, mixed molten-salt ($Na_2WO_4$-$Na_2SO_4$) droplets containing Na, W, S, and O are intentionally deposited on few-layer $WS_2$ flakes. The as-printed droplets switch between the 1D crystallographic etching mode and 1D epitaxial growth mode simply by turning off and on the S supply. Therefore, precise control over the size and composition of the molten-salt droplets, reaction temperatures, and redox environments will enable the nanofabrication of intricate and well-defined 1D-2D mixed-dimensional TMDC-based nanostructures. It holds promise for the realization of densely integrated all-TMD circuits for advanced electronic and optoelectronic devices.

**Supporting Information**
Supporting Information is available from the Wiley Online Library or from the author.

**Acknowledgements**
S.L. acknowledges the support from JSPS-KAKENHI (21K04839) and International Center for Young Scientists (ICYS), NIMS. Y.-C.L and K.S. acknowledge to JSPS-KAKENHI (22H05478) and JST-CREST program (JPMJCR20B1, JPMJCR20B5, JPMJCR1993) and JSPS A3 Foresight Program. H.E.L. acknowledges the support from JSPS-KAKENHI (23K04530) and Murata Science Foundation. S.L. acknowledges all staff members of the Nanofabrication Group at NIMS for their support.

**Author contributions**
S.L. designed and conducted the growth and etching experiments. S.L., T.T., K.T., and Y.S. carried out the Raman, photoluminescence, fluorescence, and AFM. Y.-C.L., Y.C., and K.S. performed and interpreted the STEM data. Y.D., Z.N., and Z.S. performed the SHG and analyzed the data. H.N. and T.O. performed SEM and EDX studies of the NPs. J.W. illustrated the 1D growth and etching of TMDCs. S.L., Y.-C.L., Y.D., and H.E.L. wrote the paper. All the authors contributed to data analysis, scientific discussion and commented on the manuscript.

**Conflict of interest**
The authors declare no competing interests.

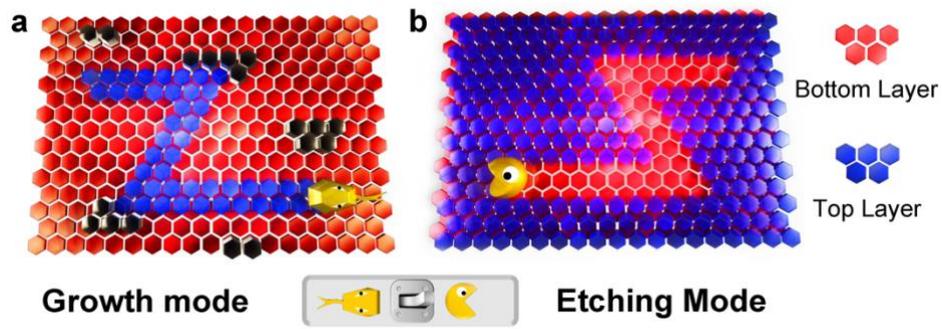

**Figure 1. The bottom-up and top-down strategies for fabricating 1D TMDC nanochannels on 2D TMDCs.**

(**a**) Schematic illustration of the bottom-up growth of 1D nanoribbon on 2D TMDC.

(**b**) Schematic illustration of the top-down etching of 2D TMDC for 1D nanotrench.

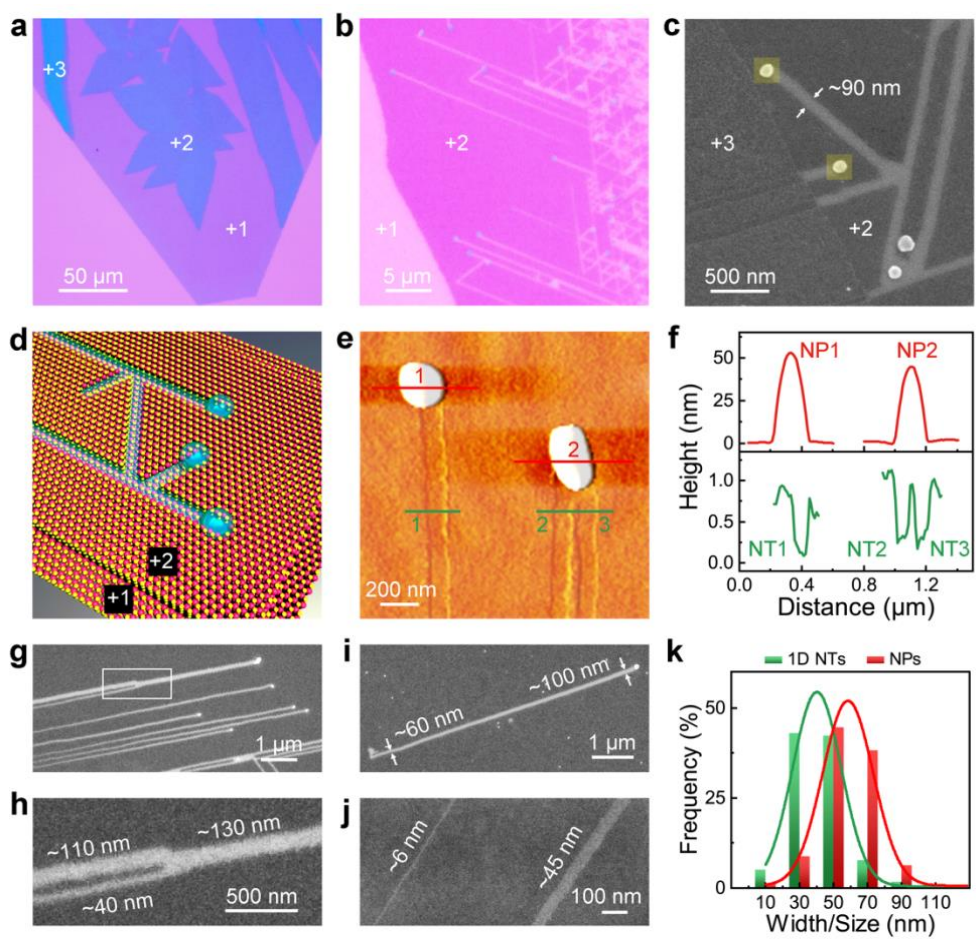

**Figure 2. 1D crystallographic etching of few-layer WS$_2$.**
(**a**) Optical image showing a molten salt CVD-grown WS$_2$ flake with monolayer (+1), bilayer (+2), and trilayer (+3) regions. (**b**) Optical image of an as-etched WS$_2$ flake with 1D nanotrenches (NTs) in the bilayer (+2) region. (**c**) SEM image showing two nanoparticles (NPs) etching the topmost WS$_2$ layer and crossing from the trilayer (+3) to the bilayer (+2) region. The NPs are highlighted with yellow boxes. (**d**) Schematic illustration of 1D NTs formed on bilayer WS$_2$ under the assistance of NPs. (**e**) AFM topography image showing as-etched WS$_2$ with three 1D NTs and two NPs. (**f**) AFM height profiles of the corresponding NPs and 1D NTs shown in (**e**). (**g**) SEM image of parallel 1D NTs on as-etched bilayer WS$_2$. (**h**) Enlarged SEM image showing the merge of two adjacent 1D NTs due to the aggregation of two NPs, outlined area in (**g**). (**i**) SEM image showing an individual NT broadening from ~60 nm to ~100 nm in width. (**j**) SEM image indicating a 1D NT with a narrow channel width of ~6 nm. (**k**) The width and size distributions of the 1D NTs and the NPs derived from the WS$_2$ flakes grown with 20 mM Na$_2$WO$_4$ aqueous solution, and the etching was conducted at 600 °C.

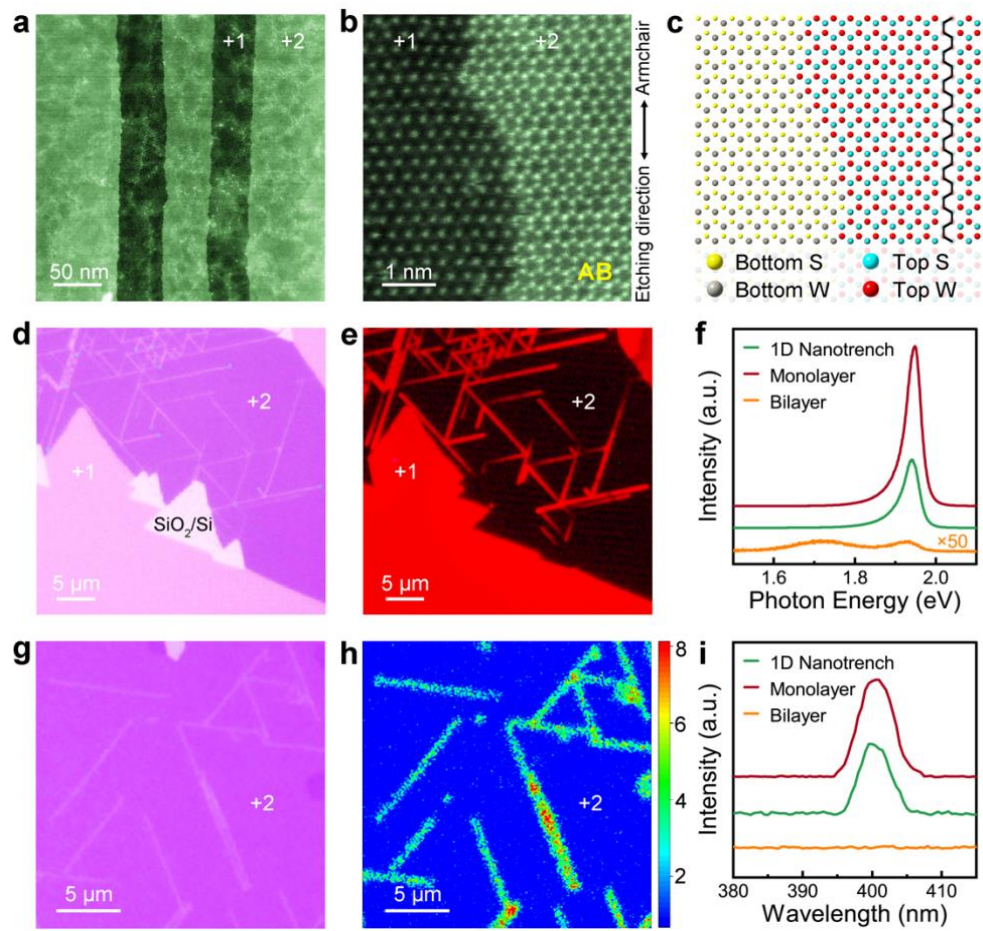

**Figure 3. Atomic structure and optical properties of as-etched $WS_2$.**
(**a**) Low-magnification STEM image showing two parallel 1D NTs in as-etched bilayer $WS_2$. (**b**) Atomic-resolution ADF-STEM image and the corresponding (**c**) atomic structure showing the edge of a 1D NT embedded in the as-etched bilayer $WS_2$. The etching takes place along the armchair direction. (**d**) Optical image and (**e**) corresponding fluorescence image of an as-etched $WS_2$ flake with 1D NTs in the bilayer region. (**f**) Typical PL spectra obtained from 1D NTs, monolayer, and bilayer regions of the as-etched $WS_2$ flakes. (**g**) Optical image and (**h**) corresponding SHG mapping results of an as-etched $WS_2$ flake. (**i**) Typical SHG spectra obtained from 1D NTs, monolayer, and bilayer regions of the as-etched $WS_2$ flakes excited with a femtosecond laser at a wavelength of 800 nm.

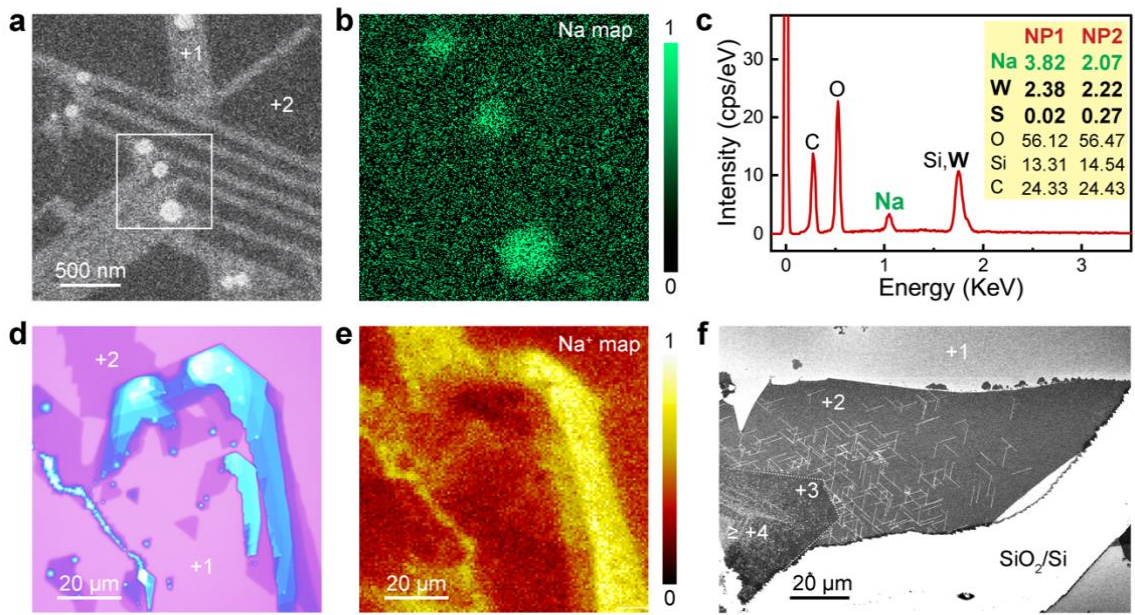

**Figure 4. Chemical composition of the nanoparticles for 1D crystallographic etching.** (**a**) SEM image of the bilayer region of an as-etched WS$_2$ flake with NPs and 1D NTs. (**b**) EDX map of the outlined area in (**a**) indicates Na accumulation in the NPs. (**c**) Typical EDX spectrum and chemical composition of the NPs. Inset is the atomic composition of two NPs. (**d**) Optical image and (**e**) corresponding TOF-SIMS map of Na$^+$ indicating the preferential accumulation of Na at the thick WS$_2$ regions (bright yellow region). (**f**) SEM image indicating the density gradient of NPs (white dots) and 1D NTs (white lines) on an as-etched WS$_2$ flake.

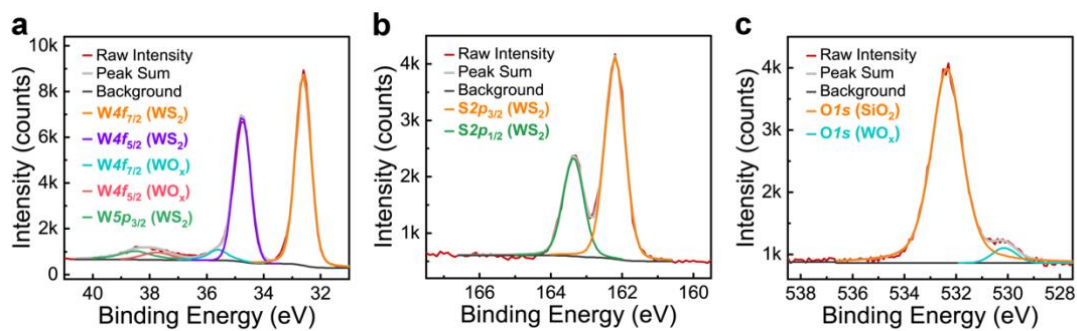

**Figure 5. XPS analysis of the molten salt CVD-grown WS₂ flakes.**

(**a**) XPS fine spectra of W$5p_{3/2}$ at 38.06 eV and W$4f$ with four fitted peaks at 32.61 eV and 34.76 eV for $WS_2$, 35.60 eV and 37.75 eV for $WO_x$. (**b**) XPS fine spectra of S$2p$ with two peaks at 163.36 eV and 162.20 eV for $WS_2$. (**c**) XPS fine spectra of O$1s$ with two separated peaks at 532.36 eV for $SiO_2$ and 530.15 eV for $WO_x$.

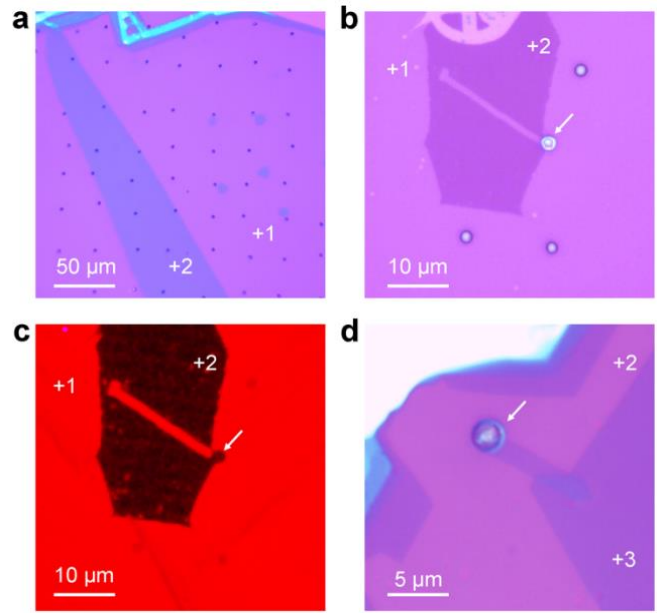

**Figure 6. Tailoring 1D crystallographic etching and growth of WS$_2$ via the sulfur switch.** (**a**) Optical image of inkjet-printed Na$_2$WO$_4$-Na$_2$SO$_4$ particles on a WS$_2$ flake. (**b**) Optical image and (**c**) corresponding fluorescence image of a 1D trench in the bilayer region of a WS$_2$ flake by activating the inkjet-printed Na$_2$WO$_4$-Na$_2$SO$_4$ particles at 700 °C in a sulfur-free atmosphere. (**d**) Optical image of a WS$_2$ ribbon grown from the inkjet-printed Na$_2$WO$_4$-Na$_2$SO$_4$ particle with sufficient sulfur supply at 800 °C.

# Supporting Information

**One-Dimensional Crystallographic Etching of Few-Layer WS$_2$**

*Shisheng Li,[1]\* Yung-Chang Lin,[2,3] Yiling Chiew,[3] Yunyun Dai,[4] Zixuan Ning,[4] Hideaki Nakajima,[5] Hong En Lim,[6] Jing Wu,[7] Yasuhisa Naito,[8] Toshiya Okazaki,[5] Zhipei Sun,[4] Kazu Suenaga,[2,3] Yoshiki Sakuma,[9] Kazuhito Tsukagoshi[1] and Takaaki Taniguchi[1]*

[1] Research Center for Materials Nanoarchitectonics (MANA), National Institute for Materials Science (NIMS), Tsukuba 305-0044, Japan

[2] Nanomaterials Research Institute, National Institute of Advanced Industrial Science and Technology, AIST Central 5, Tsukuba 305-8565, Japan

[3] The Institute of Scientific and Industrial Research (ISIR-SANKEN), Osaka University, Osaka 567-0047, Japan

[4] QTF Centre of Excellence, Department of Electronics and Nanoengineering, Aalto University, Espoo 02150, Finland

[5] Nano Carbon Device Research Center, National Institute of Advanced Industrial Science and Technology, AIST Central 5, Tsukuba 305-8564, Japan

[6] Department of Chemistry, Graduate School of Science and Engineering, Saitama University, Saitama 338-8570, Japan

[7] Institute of Materials Research and Engineering (IMRE), Agency for Science, Technology and Research (A*STAR), 138634, Singapore

[8] Nanosystem Research Institute, National Institute of Advanced Industrial Science and Technology (AIST), Tsukuba 305-8562, Japan

[9] Research Center for Electronic and Optical Materials, National Institute for Materials Science (NIMS), Tsukuba 305-0044, Japan

\* Correspondence and requests for materials should be addressed to
LI.Shisheng@nims.go.jp

**Materials and Methods**

Deposition of $Na_2WO_4$ on $SiO_2$/Si substrates

To grow few-layer $WS_2$ flakes, 20 and 30 mM $Na_2WO_4$ aqueous solution was spin-coated on $SiO_2$/Si substrates treated with UV-$O_3$ for 30 minutes to obtain hydrophilic surfaces. The spin-coating was performed at a speed of 6000 rpm for 30 seconds.

Molten-salt CVD growth of 2D $WS_2$ flakes

The $SiO_2$/Si substrates with $Na_2WO_4$ were loaded into the center of a 2-inch tube furnace. And ~500 mg sulfur in a quartz crucible was loaded upstream. The CVD chamber was pumped down to ~1 Pa. Then, the pressure was restored and maintained at atmospheric pressure with 100 sccm Ar-(5$_{vol.}$%) $H_2$ forming gas. Later, the tube furnace was heated to 800 ºC at a ramping rate of 15 ºC/min. Simultaneously, sulfur vapor sublimated from its melt at a temperature of ~200 ºC to initiate the sulfurization of $Na_2WO_4$ at 800 ºC for 10 minutes. Finally, the furnace was cooled down to room temperature within 30 minutes under the protection of pure Ar.

1D crystallographic etching of few-layer $WS_2$ flakes

The molten-salt CVD-grown $WS_2$ flakes on $SiO_2$/Si substrates were utilized for the etching experiments. The samples were first loaded in the 2-inch tube furnace. Then, the CVD system was pumped down to ~1 Pa. Later, it was restored to atmospheric pressure with 90 sccm Ar and 10 sccm Ar-(5$_{vol.}$%) $H_2$ forming gas. The etching experiments were conducted under atmospheric pressure at temperatures 550-700 ºC for 10 minutes without sulfur. Finally, the furnace was cooled down to room temperature within 20 minutes.

Transfer of as-etched $WS_2$ flakes

To transfer the as-etched $WS_2$ flakes for STEM, PMMA film was first spin-coated on the $SiO_2$/Si substrate. Then, the $WS_2$-embedded PMMA film was peeled from the substrate in 35 $_{wt.}$% KOH aqueous solution. After rinsing in DI water 3 times, the $WS_2$-embedded PMMA film was harvested with a TEM grid. After drying under ~20% relative humidity (RH) at room temperature for 8 hours, the PMMA film was removed by rinsing in acetone and IPA sequentially.

1D etching and growth of $WS_2$ with inkjet-printed of $Na_2WO_4$-$Na_2SO_4$ particles.

To avoid the influence of in-situ generated Na-W-S-O nanoparticles in the molten-salt CVD-grown $WS_2$ flakes. The samples are sequentially rinsed in IPA/$H_2O$ (9:1) and IPA. Then, an array of Na$_2$WO$_4$-Na$_2$SO$_4$ particles is printed on the Na-free WS$_2$ flakes using an inkjet printer (Fujifilm Dimatix 2831) equipped with 1-pL nozzles. Na$_2$WO$_4$-Na$_2$SO$_4$ (2 mM : 2 mM) aqueous solution is employed as ink. The etching experiments are conducted following the procedure described in "*1D crystallographic etching of few-layer WS$_2$ flakes*". The optimal etching temperature is 700 °C. The growth experiments follow the procedure of "*Molten-salt CVD growth of 2D WS$_2$ flakes*".

Raman and photoluminescence (PL) spectroscopies
The micro-Raman/PL was performed using a laser confocal microscope (inVia, Renishaw). The 532-nm excitation laser was focused on the sample surface with a 100× objective lens. Then, Raman/PL signals from the WS$_2$ flakes were detected by an electron-multiplying CCD detector (Andor) through a grating with 1800 grooves/mm for Raman and 300 grooves/mm for PL. The laser spot size is about 1 µm in diameter. Raman mapping was conducted on the WS$_2$ flakes with a step of 0.3 µm.

Atomic force microscopy (AFM)
The AFM observations were performed in tapping mode by using Asylum MFP-3D origin.

Scanning transmission electron microscopy (STEM)
STEM images of as-etched WS$_2$ flakes were acquired using JEOL Triple-C#3, an ARM200F-based ultrahigh vacuum microscope equipped with a JEOL dodecaple correctors and a cold field emission gun operating at 60 kV. The probe current was about 25-30 pA. The convergence semiangle was 37 mrad and the inner acquisition semiangle was 76 mrad. The EELS core loss spectrum was acquired using a Gatan Rio COMS camera optimized for low-voltage operation.

Scanning electron microscopy (SEM) – Energy dispersive X-ray spectroscopy (EDX)
SEM-EDX was conducted using a field-emission SEM (Hitachi, SU8220) with an annular-type EDX (Brucker, Quantax FlatQUAD). All SEM-EDX measurements were performed under a 5-kV acceleration voltage and 10-µA emission current. The energy resolution of EDX is ~130 eV at Mn Kα. Elemental mapping of EDX and quantitative analysis were conducted with ESPRIT analytical software. The analysis included spectral fitting and calibrations of the relative detection sensitivity for each atom.

<u>Time</u> <u>of</u> <u>flight</u> – <u>Secondary-ion</u> <u>mass</u> <u>spectrometry</u> <u>(TOF-SIMS)</u>

TOF-SIMS experiments were performed using the PHI TRIFT V, nano TOF equipped 30 keV $Bi^+$ liquid-metal primary ion source filtered for the $Bi^+$ ion. A compressed 12 ns primary ion pulse width was used, and an extraction voltage of 3085 V was applied to the sample stage. A primary ion dose of $5 \times 10^{11}$ ions/cm$^2$ over an area of $100 \times 100$ μm$^2$ was used for data acquisition.

<u>X-ray</u> <u>photoelectron</u> <u>spectroscopy</u> <u>(XPS)</u>

To investigate the chemical composition and chemical bonding states of the $WS_2$ samples, X-ray photoelectron spectroscopy (XPS) is performed in the PHI Quantera SXM system with a monochromatic Al Kα (1486.96 eV) source. The wide scans are performed with a Φ200 μm analysis area. The high-resolution scans are performed with a 55-eV pass energy and a 0.1-eV energy step. All spectra are calibrated by the C1s peak at 285 eV.

<u>Second-harmonic generation spectroscopy (SHG)</u>

SHG measurements are carried out with a home-built femtosecond laser-based microscopic setup. As the excitation source, we use a femtosecond laser (Spectra-Physics) at the wavelength of 800 nm with a repetition rate of 84 MHz. The average power is ~30 mW with the pulse duration of ~60 fs, which yields an estimated pulse peak power of ~6 kW. The laser beam is focused on the samples through an objective lens (60×, NA = 0.5). The nonlinear optical signals are detected using a photomultiplier tube (PMT) following a monochromator (Andor 328i) in the reflection configuration. For SHG mapping, the sample is raster scanned by the laser beam with a step size of ~50 nm.

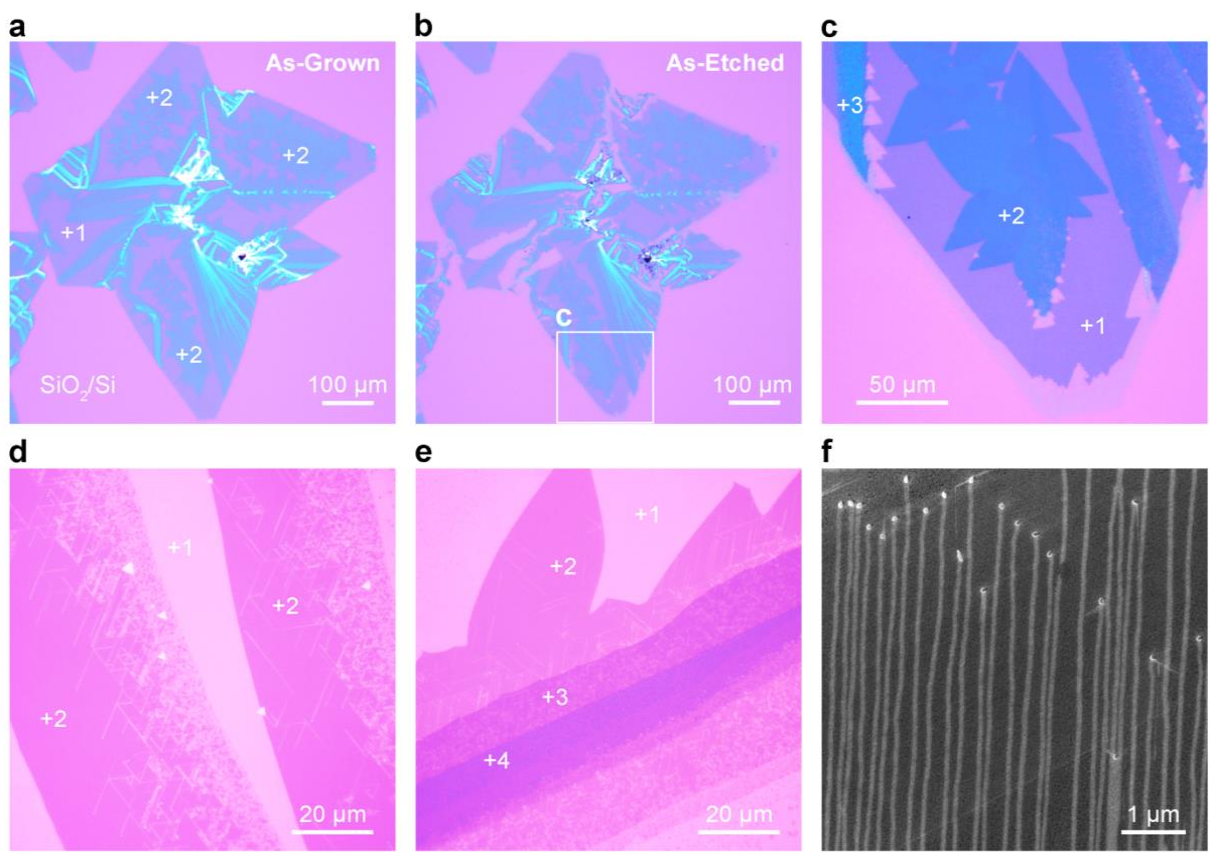

**Supplementary Figure 1. Morphologies of as-grown and as-etched WS$_2$ flakes.** (**a**) Optical image of as-grown WS$_2$ flakes on a SiO$_2$/Si substrate. (**b**, **c**) Optical images of the same WS$_2$ flake after etching. (**d**-**e**) Optical images show typical 1D NTs in the bilayer, trilayer, and thick-layer regions of WS$_2$ flakes. Here, +1, +2, +3, and +4 labels correspond to the monolayer, bilayer, trilayer, and thick-layer regions of WS$_2$ flakes. (**f**) SEM image of parallel 1D NTs in the bilayer region of a WS$_2$ flake forming a comb-like structure.

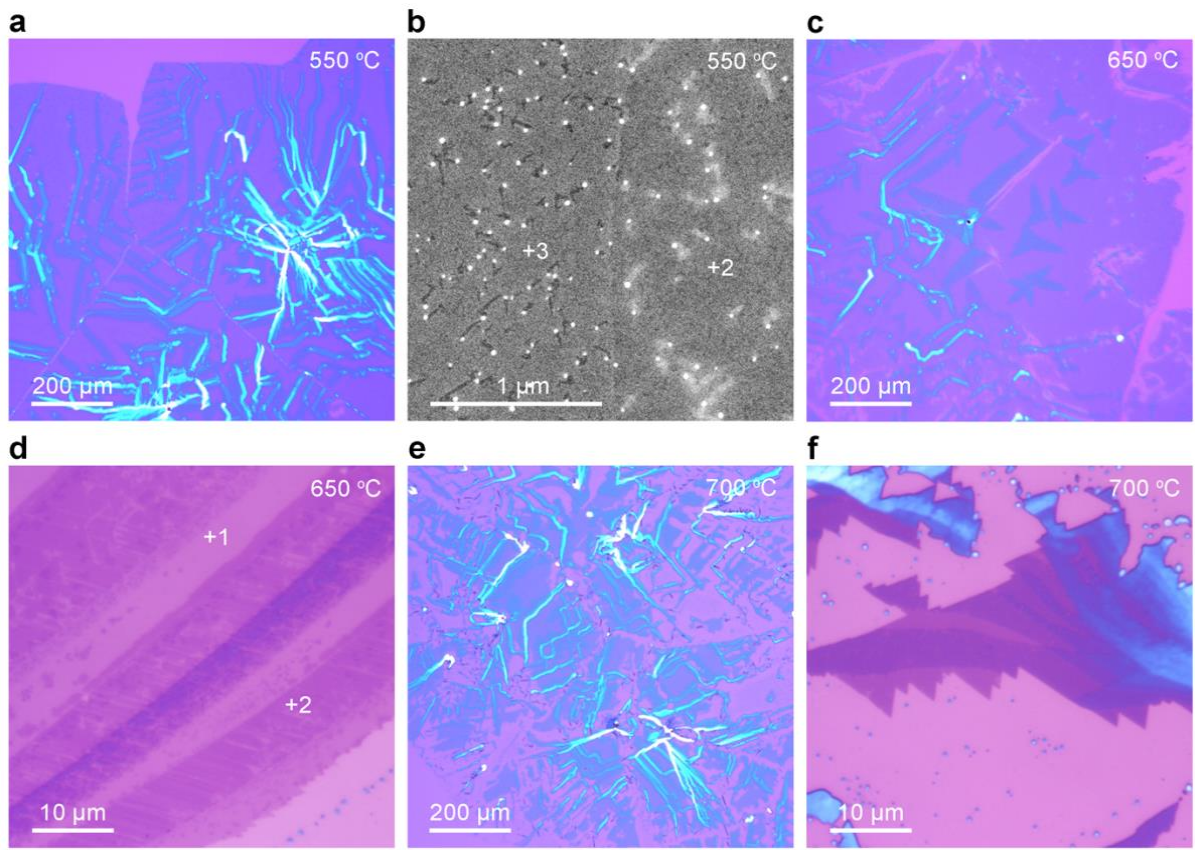

**Supplementary Figure 2. Temperature-dependent etching of molten salt CVD-grown WS$_2$ flakes.** (**a**) Optical and (**b**) SEM images of WS$_2$ flakes after etching at 550 °C under Ar-(0.5$_{vol.}$%) H$_2$. The optical image indicates no obvious morphological change. The high-resolution SEM image demonstrates the NPs (bright dots) and short 1D NTs in the as-etched WS$_2$ flake (dark lines in the trilayer region and bright lines in the bilayer region). (**c-f**) Optical images of WS$_2$ flakes after etching at (**c, d**) 650 °C, and (**e, f**) 700 °C under Ar-(0.5$_{vol.}$%) H$_2$.

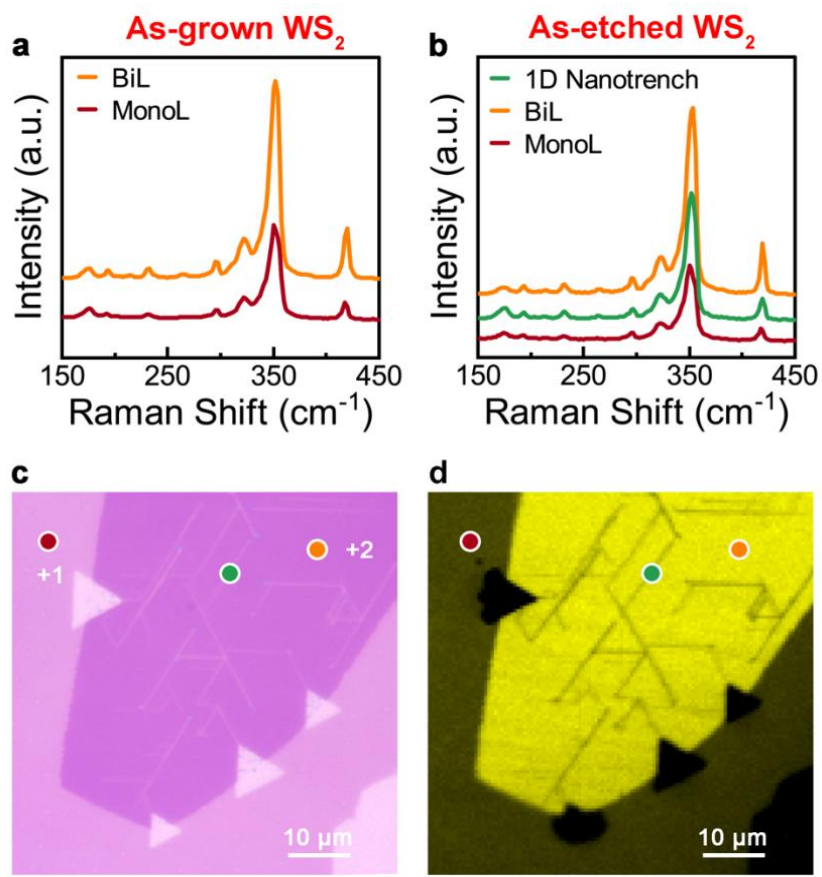

**Supplementary Figure 3. Raman characterization of as-grown and as-etched WS$_2$ flakes.** (**a**) Raman spectra of as-grown WS$_2$ flakes. (**b**) Raman spectra of as-etched WS$_2$ flakes. All the spectra represent the average of 10-20 spectra collected from different locations of the WS$_2$ flakes. (**c**) Optical image and (**d**) corresponding Raman intensity map of 2LA (352 cm$^{-1}$).

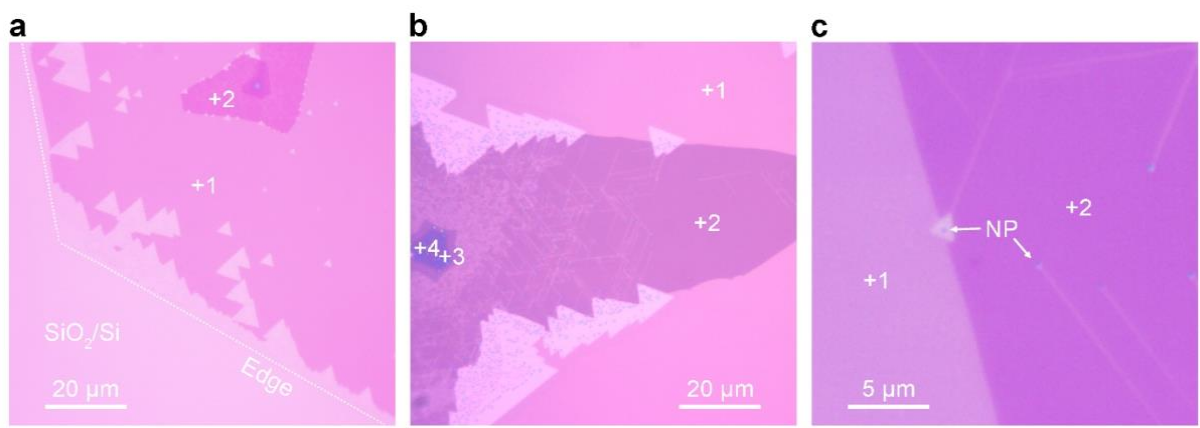

**Supplementary Figure 4. The layer number-dependent 1D and 2D etching of WS$_2$.** (**a**) Optical image demonstrating 2D etching-induced triangle holes in the monolayer region of a WS$_2$ flake. (**b**) Optical image showing triangle holes at the boundary of monolayer-bilayer WS$_2$. (**c**) Optical image of an NP etching from the bilayer region and terminating at the monolayer-bilayer boundary with a triangle hole. It indicates a transition from 1D to 2D etching.

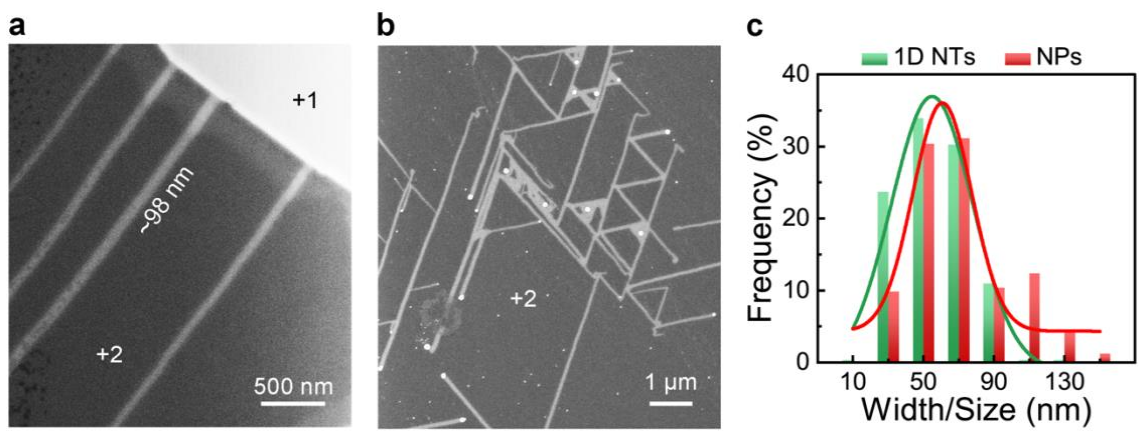

**Supplementary Figure 5. The concentration-dependent 1D NT width and NP size distribution.** (**a**) SEM image shows a 1D NT with large width of ~98 nm in the as-etched WS$_2$ flake grown with 20 mM Na$_2$WO$_4$. (**b**) Typical SEM image and (**c**) 1D NT width and NP size distribution of the as-etched WS$_2$ flakes grown with 30 mM Na$_2$WO$_4$.

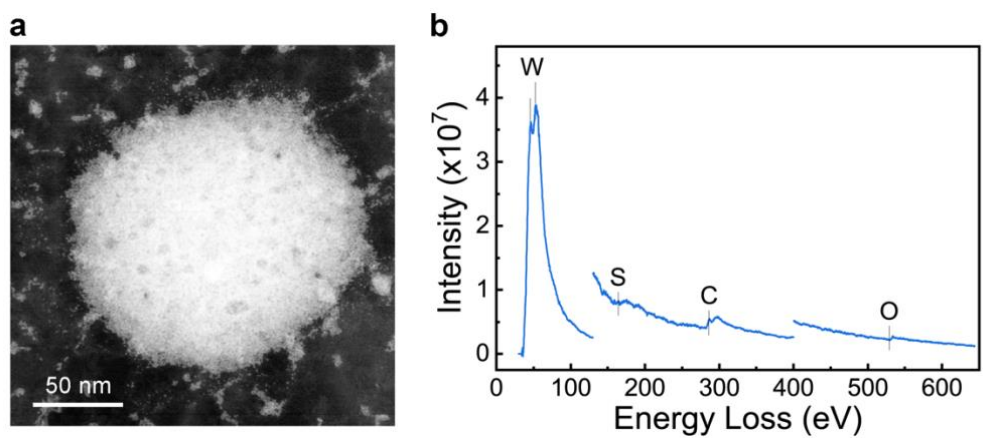

**Supplementary Figure 6. EELS spectra of a typical NP on etched WS$_2$ flake.** (**a**) STEM image of a NP on an etched WS$_2$ flake. (**b**) EELS spectra obtained from the NP indicating the existence of W, S and O, C signal is from the ambient. The Na is dissolved in H$_2$O during the wet-transfer process for TEM sample preparation.

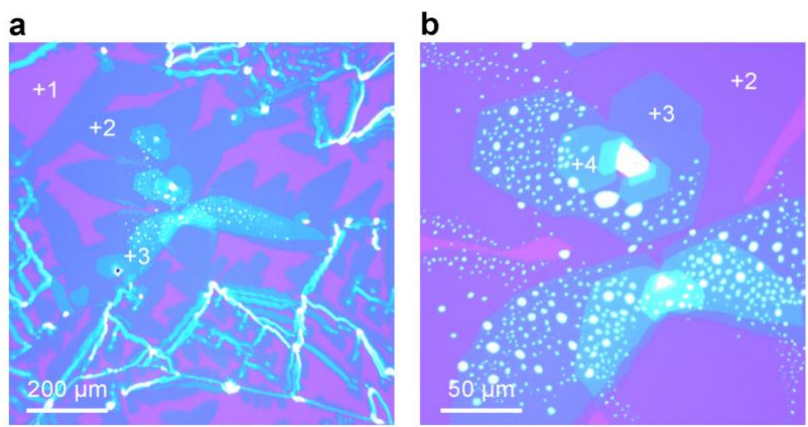

**Supplementary Figure 7. The layer number-dependent density gradient of intermediate Na-W-S-O melts on as-grown WS$_2$ flakes.** (**a**, **b**) Optical images of molten salt CVD-grown WS$_2$ flakes with residual Na-W-S-O melts in the thick-layer regions. The growth time is shortened to see the distribution of the Na-W-S-O melts.

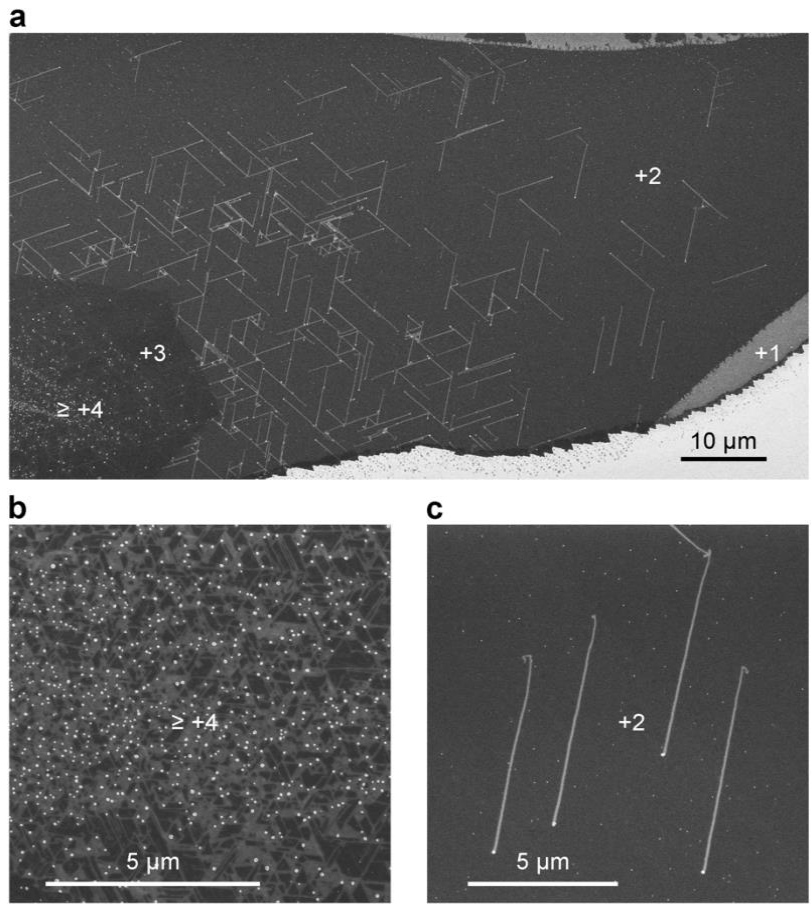

**Supplementary Figure 8. The layer number-dependent distribution of NPs and 1D NTs in as-etched WS$_2$.** (**a**) SEM image of an as-etched WS$_2$ flake with regions of different layer numbers. (**b**) SEM image demonstrates the high-density NPs and 1D NTs on the surface of the thick-layer (≥ +4) region. (**c**) SEM image demonstrates the sparse distribution of NPs and 1D NTs in the bilayer region far from the thick-layer region.
10

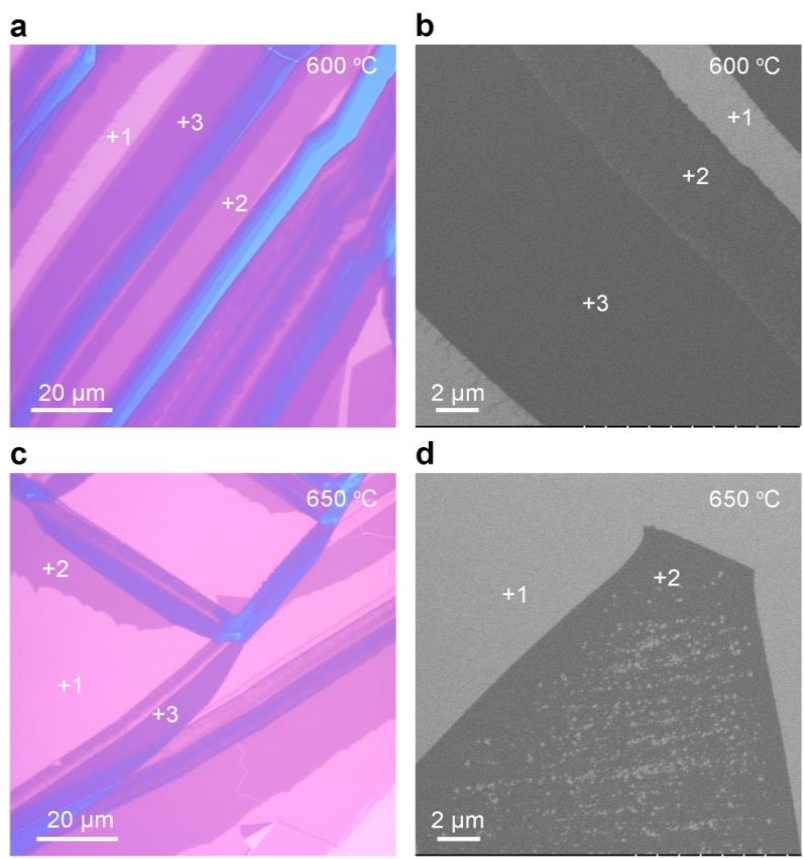

**Supplementary Figure 9. Sodium-free etching of transferred WS$_2$ flakes on SiO$_2$/Si substrates**. (**a**) Optical and (**b**) SEM images of WS$_2$ flakes etched at 600 °C under Ar-(0.5$_{vol.}$%) H$_2$. No obvious morphological change is observed. (**c**) Optical and (**d**) SEM images of the WS$_2$ flakes etched at 650 °C under Ar-(0.5$_{vol.}$%) H$_2$. Only small triangle etching holes are observed in the bilayer region.

**Note:** To remove the Na residual, the as-grown WS$_2$ flakes are first detached from growth substrates and floating on the surface of DI water. Then, UV-O$_3$-treated SiO$_2$/Si substrates with hydrophilic surfaces are used to harvest the floating WS$_2$ flakes. The etching experiments are conducted at 600 and 650 °C under Ar-(0.5 $_{vol.}$%) H$_2$.

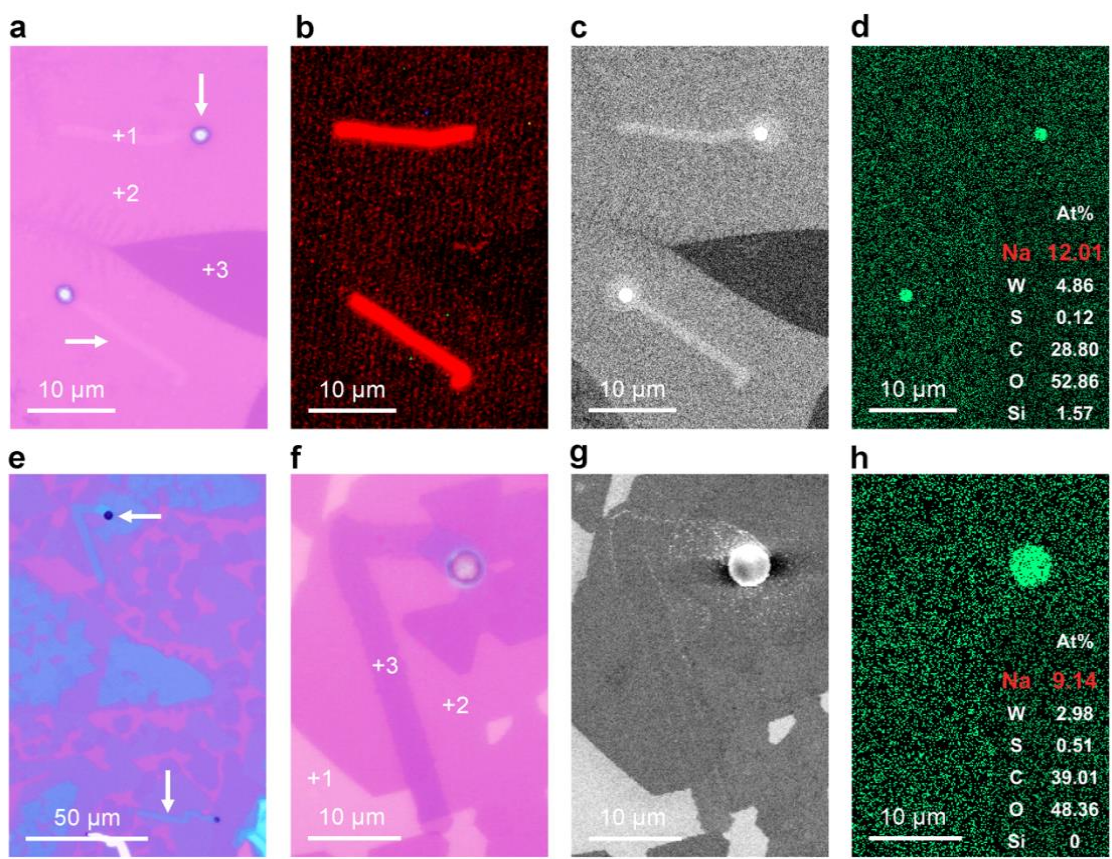

**Supplementary Figure 10. 1D crystallographic etching and 1D epitaxial growth using inkjet-printed Na$_2$WO$_4$-Na$_2$SO$_4$ particles.** (**a**) Optical and (**b**) corresponding fluorescence images of two 1D WS$_2$ trenches in the bilayer region of a WS$_2$ flake after etching at 700 °C under Ar-(0.5 $_{vol.}$%) H$_2$ without sulfur supply. (**c**) SEM image and (**d**) corresponding Na EDX map of as-etched WS$_2$ flakes. The bright spots in (**d**) indicate the accumulation of Na in the particles. The inset of (**d**) displays the atomic composition estimated from the EDX spectrum of a particle. (**e**, **f**) Optical images show the 1D WS$_2$ ribbons grown on top of the bottom WS$_2$ layers at 800 °C in a sulfur-sufficient condition. (**g**, **h**) SEM image and corresponding Na EDX map of as-grown 1D WS$_2$ ribbon. The bright spots in (**h**) indicate the accumulation of Na in the particle. The inset of (**h**) displays the atomic composition calculated from the EDX spectrum of the particle.

ToC figure

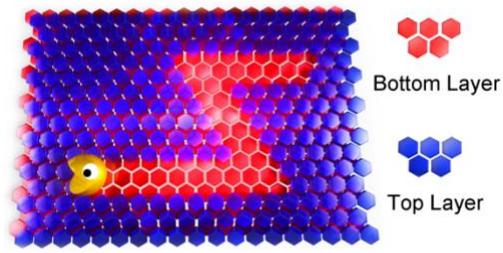

A novel technique using the Na-W-S-O droplets to create one-dimensional nanotrenches in few-layer $WS_2$, resulting in enhanced photoluminescence and second harmonic generation.